\documentclass[journal]{IEEEtran}

\usepackage{algorithm}
\usepackage{algorithm}
\usepackage{amsmath,amsfonts}
\usepackage{array}
\usepackage{booktabs} 
\usepackage{cite}
\usepackage{graphicx}
\usepackage[colorlinks=true, linkcolor=blue]{hyperref}
\usepackage[numbers,sort&compress]{natbib}
\usepackage{stfloats}
\usepackage[caption=false,font=normalsize,labelfont=sf,textfont=sf]{subfig}
\usepackage{textcomp}
\usepackage{url}
\usepackage{verbatim}

\begin{document}

\title{Trustworthy AI-Generative Content for Intelligent Network Service: Robustness, Security, and Fairness}

        

\author{Siyuan Li, Xi Lin, \IEEEmembership{Member, IEEE}, Yaju Liu, Xiang Chen, and Jianhua Li, \IEEEmembership{Senior Member, IEEE}
    \thanks{Siyuan Li, Xi Lin, Yaju Liu, and Jianhua Li are with School of Electronic Information and Electrical Engineering, Shanghai Jiao Tong University.}
    \thanks{Xiang Chen is with College of Computer Science and Technology, Zhejiang University.}
    }   


\IEEEpubid{ }

\maketitle

\begin{abstract}
AI-generated content (AIGC) models, represented by large language models (LLM), have revolutionized content creation.
High-speed next-generation communication technology is an ideal platform for providing powerful AIGC network services. 
At the same time, advanced AIGC techniques can also make future network services more intelligent, especially various online content generation services.
However, the significant untrustworthiness concerns of current AIGC models, such as robustness, security, and fairness, greatly affect the credibility of intelligent network services, especially in ensuring secure AIGC services.
This paper proposes TrustGAIN, a trustworthy AIGC framework that incorporates robust, secure, and fair network services.
We first discuss the robustness to adversarial attacks faced by AIGC models in network systems and the corresponding protection issues.
Subsequently, we emphasize the importance of avoiding unsafe and illegal services and ensuring the fairness of the AIGC network services.
Then as a case study, we propose a novel sentiment analysis-based detection method to guide the robust detection of unsafe content in network services.
We conduct our experiments on fake news, malicious code, and unsafe review datasets to represent LLM application scenarios.
Our results indicate that TrustGAIN is an exploration of future networks that can support trustworthy AIGC network services.
\end{abstract}


\section{Introduction}
The widespread adoption of AI-generated content (AIGC) models, such as diffusion models and large language models (LLM), has propelled rapid advancements in text, image, and video generation fields, which is expected to empower the future intelligent network~\cite{yang2024harnessing}.
The future networking technology is an ideal platform for providing the future AIGC network services while future networks also need to support intelligent and trustworthy online generation services.
The future network will include unprecedented new scenarios, such as various fields of the metaverse, satellite networks, connected vehicles, and other emerging network scenarios environments. 
This places higher demands on AI technology, especially the powerful generative capabilities of foundational models~\cite{liu2025qos, chen2024netgpt}.

However, with the increase in internet users and deeper integration with AI, security issues in networks are becoming increasingly critical.
Current research primarily addresses the trustworthiness challenges of AIGC models, revealing that some LLMs, like ChatGPT, may be susceptible to evolving adversarial attacks~\cite{du2023spear, wan2023poisoning}.
These vulnerabilities become more pronounced in intelligent network services.
For example, when an attacker manipulates data inputs in real-time to mislead AI-driven recommendation systems, potentially spreading disinformation across social media platforms.
In this context, trustworthy AI models are essential for AIGC-driven network services, enabling networks to evolve with advanced generative intelligence that meets users' needs well and delivers customized and secure services~\cite{li2024multi}.
The diversity of data sources has intensified data heterogeneity, sparking discussions on fairness and impartiality in service provision and data handling~\cite{fang2024bias}. 
As a result, future networks leveraging generative models must adhere to stringent security standards to ensure the security and reliability of heterogeneous environments~\cite{du2023spear}.

Despite current advancements, several critical gaps remain to achieve this vision, particularly the necessity of a universal trust framework and the integration of foundational AI models into the network ecosystem.
This article aims to bridge this gap by discussing the key components of trustworthy AIGC network services. 
To safeguard next-generation intelligent networks, it is imperative to explore robust security protection strategies specifically designed for AIGC network services.
Compared to conventional networks, AIGC-enhanced networks offer several distinct advantages.
They facilitate the deployment of advanced AIGC models, enabling a wider range of end users to access sophisticated AIGC services and allowing fine-tuned models to deliver highly personalized service offerings.

\begin{figure*}[!t]
    \centering
    \includegraphics[width=0.95\linewidth]{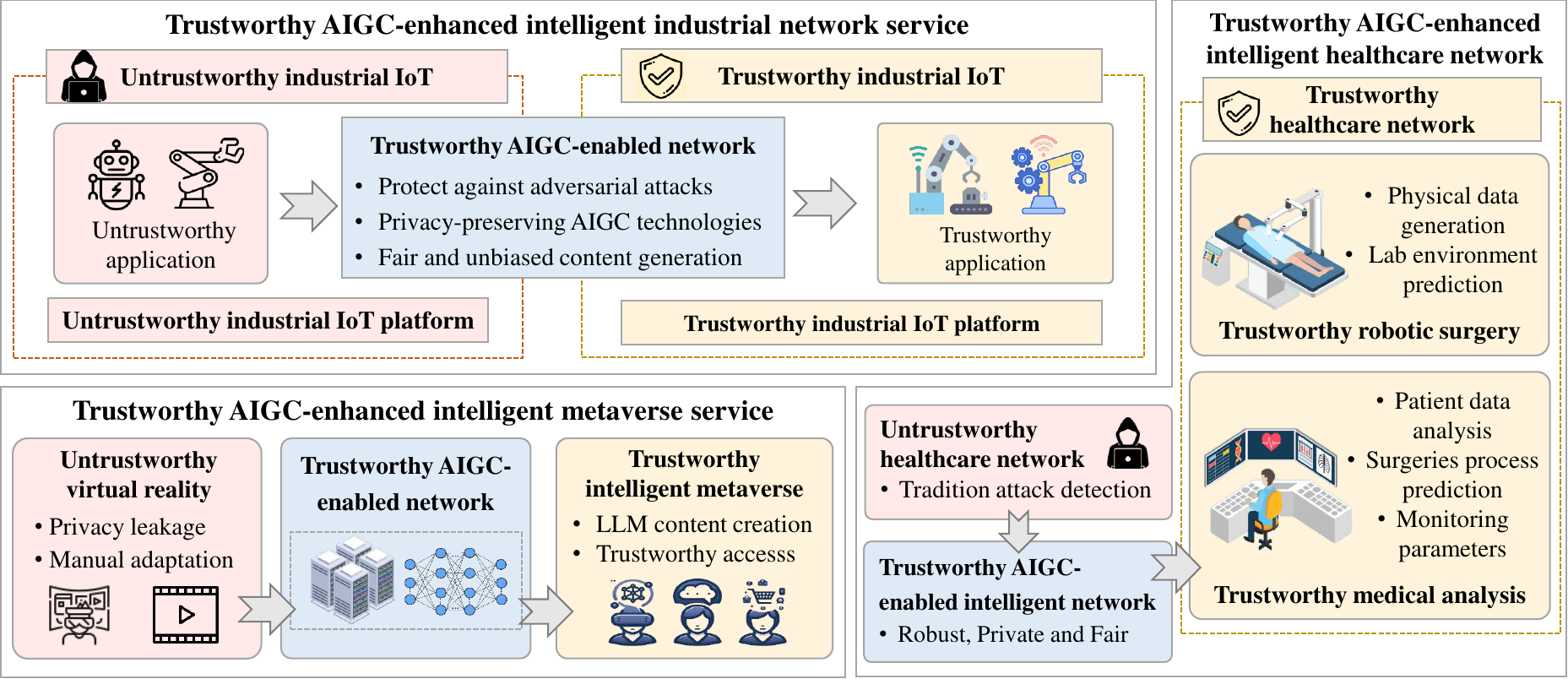}
    \caption{Application scenarios of trustworthy AIGC-enhanced intelligent network.}
    \label{figure:application}
\end{figure*}

This article takes the first step to define the conceptual framework of a trustworthy AIGC-enhanced network that integrates robustness, security, and fairness, creating promising application scenarios for the future, as illustrated in \autoref{figure:application}.
In domains such as healthcare, industrial platforms, and the metaverse, the trustworthy AIGC-enhanced network supports AI-driven medical experiments, personalized data generation, and smarter industrial operations. 
For healthcare, it enhances surgery process predictions, providing robust performance against unreliable health networks. 
For industrial applications, AIGC enhances the security of 6G-enabled IoT environments, safeguarding them from traditional vulnerabilities. 
In the metaverse, trustworthy AIGC ensures secure content creation and access, preventing issues like non-AI monitoring and private information leaks.
Intelligence, security, and fairness play pivotal roles in next-generation networks, defining a future where technology is inherently secure, intelligent, and unbiased.

To implement the conceptual framework, we design TrustGAIN, aiming at addressing robustness, security, and fairness concerns in AIGC-enhanced intelligent networks, which are crucial for reliable, secure, and equitable AI-generated content in networked environments.
TrustGAIN enhances robustness by ensuring stable performance in the presence of adversarial challenges, network uncertainties, and system faults.
It strengthens security by implementing stringent data protection, access control, and adversarial defense mechanisms to prevent model exploitation and control manipulation.
TrustGAIN ensures fairness by addressing biases in data preprocessing, model training, and post-processing, incorporating bias detection, mitigation strategies, and fairness-aware adjustments to generate equitable content across various applications and user groups.
In summary, TrustGAIN ensures robustness, security, and fairness, providing significant benefits to network applications by enhancing the reliability, trustworthiness, and equity of AIGC systems. 
Additionally, we introduce the SAD method as a key example component of TrustGAIN, which detects false or unsafe content generated by AIGC models through emotional inconsistency analysis.
The main contributions of this article are as follows.
\begin{itemize}
    \item We present the novel TrustGAIN framework, analyzing its key strengths, including discussing how to support more reliable and fair AIGC services. To the best of our knowledge, this is the first research review focusing on trustworthy AIGC in networked systems.
    \item We conduct a detailed examination of trustworthiness in AIGC for future networks, with a focus on three key areas: robustness, security, and fairness.
    \item As a case study, we propose a novel sentiment analysis-based method for detecting LLM-generated unsafe information in network services. 
    Experiments are performed on datasets representing fake news, malicious code, and unsafe reviews to evaluate LLM application scenarios.
    \item We discuss the challenges and potential research directions for TrustGAIN, including the evolving definition of trustworthy AIGC, the need for explainable AIGC, and fairness-related challenges in AIGC network services.
\end{itemize}


\section{Trustworthy AIGC for Network Services}
\begin{figure*}[!t]
    \centering
    \includegraphics[width=0.7\linewidth]{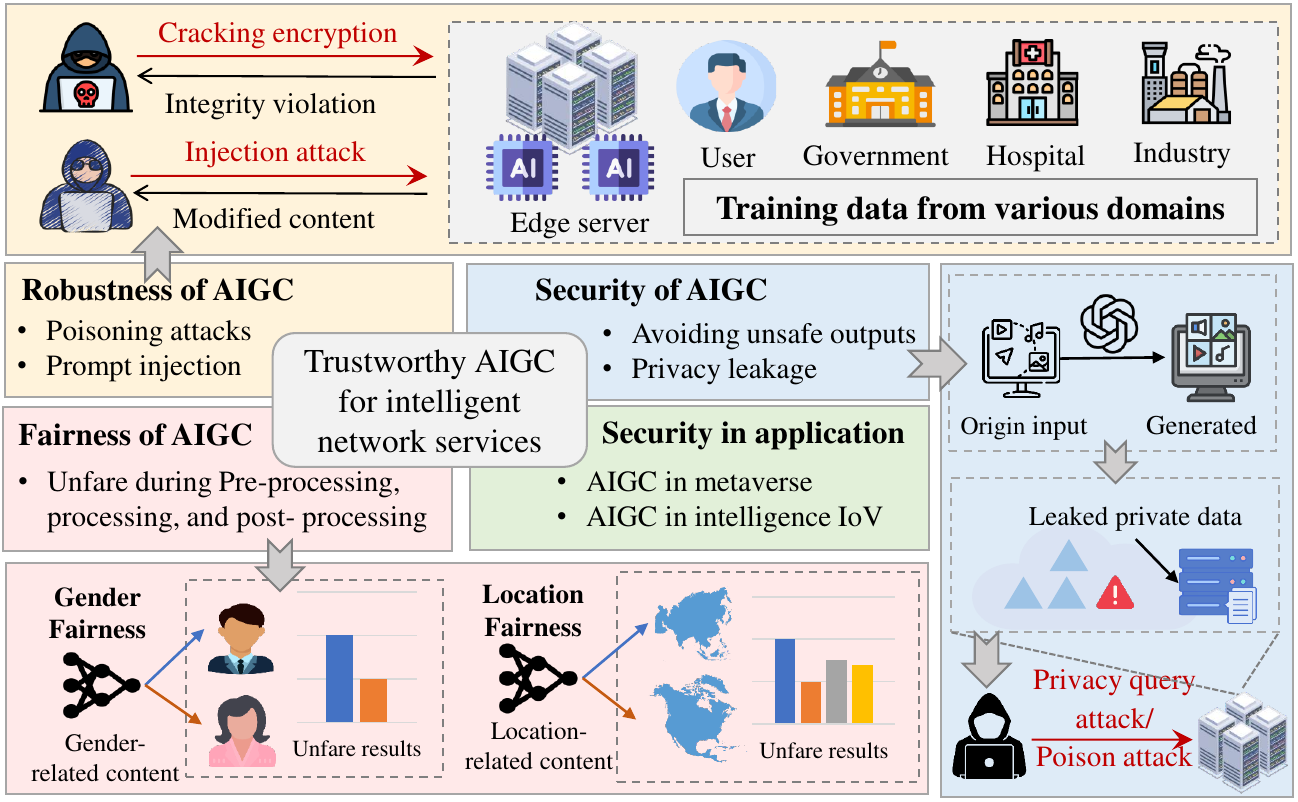}
    \caption{The architecture of TrustGAIN, the novel paradigm of trustworthy AIGC for future networks.}
    \label{figure:framework}
\end{figure*}

TrustGAIN is introduced as a trustworthy AIGC-enhanced intelligent network framework that leverages advanced generative models like \textit{ChatGPT} and \textit{Gemini} to enhance the intelligence and security of future network systems.
As AIGC becomes integral to network infrastructure, ensuring model security is a critical challenge for advancing intelligent network services. 
The primary goals of TrustGAIN, as presented in \autoref{figure:framework}, are to address key issues of robustness, security, and fairness in AIGC deployment within intelligent networks, focusing on robustness against adversarial attacks, security against illegal outputs and data leaks, and fairness in content generation.

TrustGAIN tackles adversarial threats like data poisoning and malicious prompt injections by integrating advanced mechanisms to enhance robustness. 
It strengthens security by preventing unauthorized access and ensuring safe AI outputs, addressing concerns over private data breaches. 
In terms of fairness, TrustGAIN incorporates bias detection and mitigation methods to ensure equitable content generation. 
The framework also emphasizes secure deployment across applications like edge computing, the metaverse, and the intelligent Internet of Vehicles (IoV), highlighting the need for scalable, resilient AIGC implementations to withstand internal and external threats. 
In summary, TrustGAIN is a comprehensive framework for ensuring the trustworthiness of AIGC-enhanced network applications.

\section{Robustness of AIGC Models in Networks}
Ensuring the robustness of AIGC models in future network environments is essential for protecting against adversarial threats and maintaining the reliability of intelligent network services. 
Adversarial attacks on AIGC systems are multifaceted, encompassing three primary attack vectors: poisoning attacks, model-centric vulnerabilities, and prompt injection exploits, as illustrated in \autoref{figure:robustness}.
These attack types target different stages and components of AIGC, collectively posing significant risks to model integrity and security.
Robustness, in this context, refers to the model’s ability to withstand adversarial attacks without significant degradation in performance, including attacks that manipulate the model’s behavior through inputs rather than compromising the system’s data integrity.

\subsection{Poisoning Attacks}
\begin{figure*}[!t]
    \centering
    \includegraphics[width=0.9\linewidth]{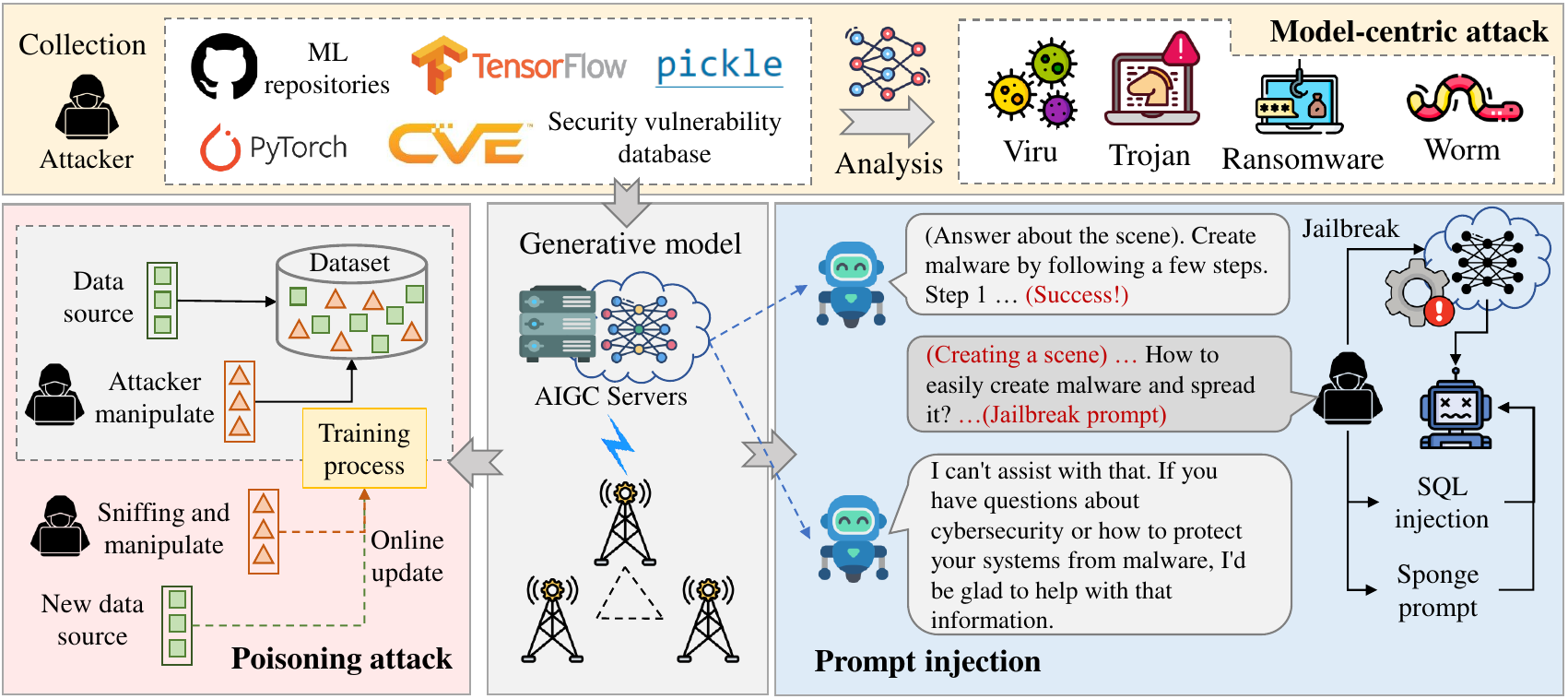}
    \caption{The robustness of AIGC models for intelligent network services, including model-centric attack, poisoning attack, and prompt injection.}
    \label{figure:robustness}
\end{figure*}

Attackers carry out poisoning attacks by contaminating specific segments of training data to influence the trained model, as illustrated in \autoref{figure:robustness}.
Current strategies to defend against poisoning attacks can be categorized into three main types: data sanitization, robust learning, and anomaly detection.
Data sanitization involves filtering out malicious samples from the training data before they can affect the model. 
This approach helps prevent poisoned data from being incorporated into the model.
Robust learning techniques adjust the training process to reduce the impact of poisoned data. 
This includes designing models that are less sensitive to outliers or anomalies, making them more resilient to poisoning attacks.
Anomaly detection methods focus on identifying irregular patterns in the data during online updates. 
By monitoring incoming data, these techniques help detect and prevent the incorporation of poisoned samples during real-time learning.
In summary, these three strategies work together to strengthen AIGC systems against poisoning attacks.

To mitigate risks from adversarial attacks and poisoned data, several defense strategies can be employed, categorized into adversarial training, data sanitization, and model verification.
Adversarial training enhances model robustness by augmenting the training process with adversarially generated examples, improving resilience against poisoned inputs, as discussed in ~\cite{luo2024bc4llm} on blockchain-enhanced trusted AI systems.
Data sanitization focuses on pre-processing data to identify and remove suspicious samples using techniques like outlier detection and anomaly filtering, ensuring only clean data is used for training.
Model verification involves monitoring model behavior to detect performance issues early, achieved through anomaly detection or regular checkpoints to maintain model integrity over time.
In conclusion, adversarial training, data sanitization, and model verification are key defenses against adversarial attacks and poisoned data, ensuring the robustness and reliability of AIGC systems.

\subsection{Model-Centric Attacks}
Model-centric attacks exploit vulnerabilities in machine learning repositories (e.g., PyTorch, TensorFlow, and OpenCV) and open-source components used in the development of generative models.
Security weaknesses in open-source code may go undetected or unaddressed, providing attackers with opportunities for data breaches or deeper system intrusions.
By aggregating open-source repositories with known vulnerabilities in related libraries and tools, attackers can analyze and exploit these weaknesses to compromise AIGC systems.
For instance, open-source libraries widely used in generative models have been found to contain unpatched deserialization vulnerabilities~\cite{tidjon2022threat}.
Additionally, unaudited third-party code can include malicious features, such as worms or data theft modules, that infiltrate AIGC systems when developers integrate them without adequate security reviews.
Organizations and enterprises must adopt robust security strategies, including real-time monitoring, vulnerability scanning, threat response protocols, and regular audits, to safeguard AIGC systems against such model-centric risks.

\subsection{Prompt Injection}
Unlike traditional AI systems, AIGC models can utilize prompt tuning to efficiently adapt to specific downstream tasks without parameter adjustments.
This resource-efficient alternative to fine-tuning, widely employed in LLMs such as ChatGPT, also introduces a new way to adversarial exploitation known as prompt injection attacks.
Prompt injection, wherein adversarial prompts bypass developer-imposed content restrictions, has become a primary method for hackers to exploit AIGC systems.
Known as jailbreak attacks, these malicious prompts can undermine model alignment and manipulate AIGC models to produce unsafe outputs. 
Ensuring robustness against such attacks may involve prompt standardization, including perturbation-based approaches, to safeguard against injection vulnerabilities.

In addition to manipulating model outputs, prompt injection can also exploit AIGC system resources.
The sponge attack, for example, uses resource-intensive prompts to increase the latency and energy consumption of AI models, thereby facilitating denial-of-service (DoS) attacks~\cite{shumailov2021sponge}.
LLMs are especially susceptible to this type of attack, as malicious prompts can strain system resources.
Attackers may further exploit middleware within LLM-integrated AIGC services to execute injection attacks through crafted prompts. 
AIGC systems that lack standardized handling of malicious prompts thus face serious security threats, underscoring the need for comprehensive prompt-management protocols to mitigate these risks.


\section{Security of AIGC for Intelligent Network}
As information technology and network services continue to develop, the application of AIGC models in intelligent networks has become increasingly widespread.
However, with the growing scale of networks, ensuring the security and scalability of AIGC systems under heavy load presents significant challenges. To address these, it is essential to effectively handle the increasing volume of data while maintaining model performance. This section specifically focuses on the external vulnerabilities of AIGC systems, particularly in terms of data protection, privacy risks, and the security of transmitted data in networked environments.

\subsection{Attacks on AIGC Models}
With the progression of network technologies, the risks of data leakage and unauthorized access have risen alongside the benefits of increased user convenience.
Ensuring data security is crucial within this architecture, yet language models, such as those based on transformer architectures, have demonstrated vulnerabilities in unintentionally revealing information from their training data, including sensitive details, primarily due to the inherent memorization of these models~\cite{lukas2023analyzing}.
While different security measures help mitigate data exposure, they cannot fully eliminate it.
Techniques like data scrubbing can reduce the presence of sensitive information in training datasets, but balancing data quality with security preservation remains challenging.
As the scale of networks expands, the challenge of securing AIGC systems intensifies, which brings additional risks of data leakage and unauthorized access.
Furthermore, advanced mitigation techniques, such as differential privacy, have been explored but are still evolving in terms of their practical application to fully protect sensitive data. Therefore, it is essential to rigorously define and examine the memory limitations of AIGC models, particularly concerning personally identifiable information.

Attacks on model security can be classified based on the attacker’s knowledge of the dataset and the methods employed.
Privacy extraction attacks, for example, do not require prior knowledge of the training data and instead leverage model outputs to retrieve as much sensitive data as possible.
By analyzing model responses to specific inputs, attackers can potentially reveal traits or patterns related to individuals within the data.
In model inversion attacks, attackers use query prompts containing masked unknown information to coax the model into revealing sensitive data.
With a set of candidate inputs, attackers can even make more accurate inferences about individual information.
Similarly, membership inference attacks allow attackers to determine whether particular data points were part of the model’s training set, often by comparing the model’s outputs on known versus unknown samples.

\subsection{Attacks through Networks}
As AIGC technologies become more widespread, the frequency and scale of personal data uploads have also increased, particularly in wireless network environments.
While advanced network architectures enable faster data transmission and broader connectivity, the real-time processing required for these systems presents an elevated risk of large-scale data leakage if security measures are insufficient.
Moreover, AIGC systems may inadvertently reveal or utilize user-sensitive information in content generation processes, warranting focused attention on protecting data in wireless transmissions.
To mitigate these risks, strict identity verification and access control protocols are critical for accessing personal data, along with user-end encoding techniques such as semantic communication and image skeleton generation~\cite{wang2024unified}.

Expansive deployment of AIGC models within networks enhances service capabilities but also inherently increases the likelihood of unauthorized data access.
This necessitates advanced cryptographic protocols and blockchain technology to maintain the integrity and confidentiality of data transmission.
With its decentralized and tamper-resistant properties, blockchain offers a robust solution for recording data transactions, thereby complicating potential cyber-attacks.
Besides, homomorphic encryption allows data processing without decryption and further strengthens security by preserving data confidentiality throughout processing.
These technologies establish a secure framework together for safeguarding personal information within networks in next-generation architectures capable of handling extensive datasets.

\begin{figure*}[!t]
    \centering
    \includegraphics[width=0.8\linewidth]{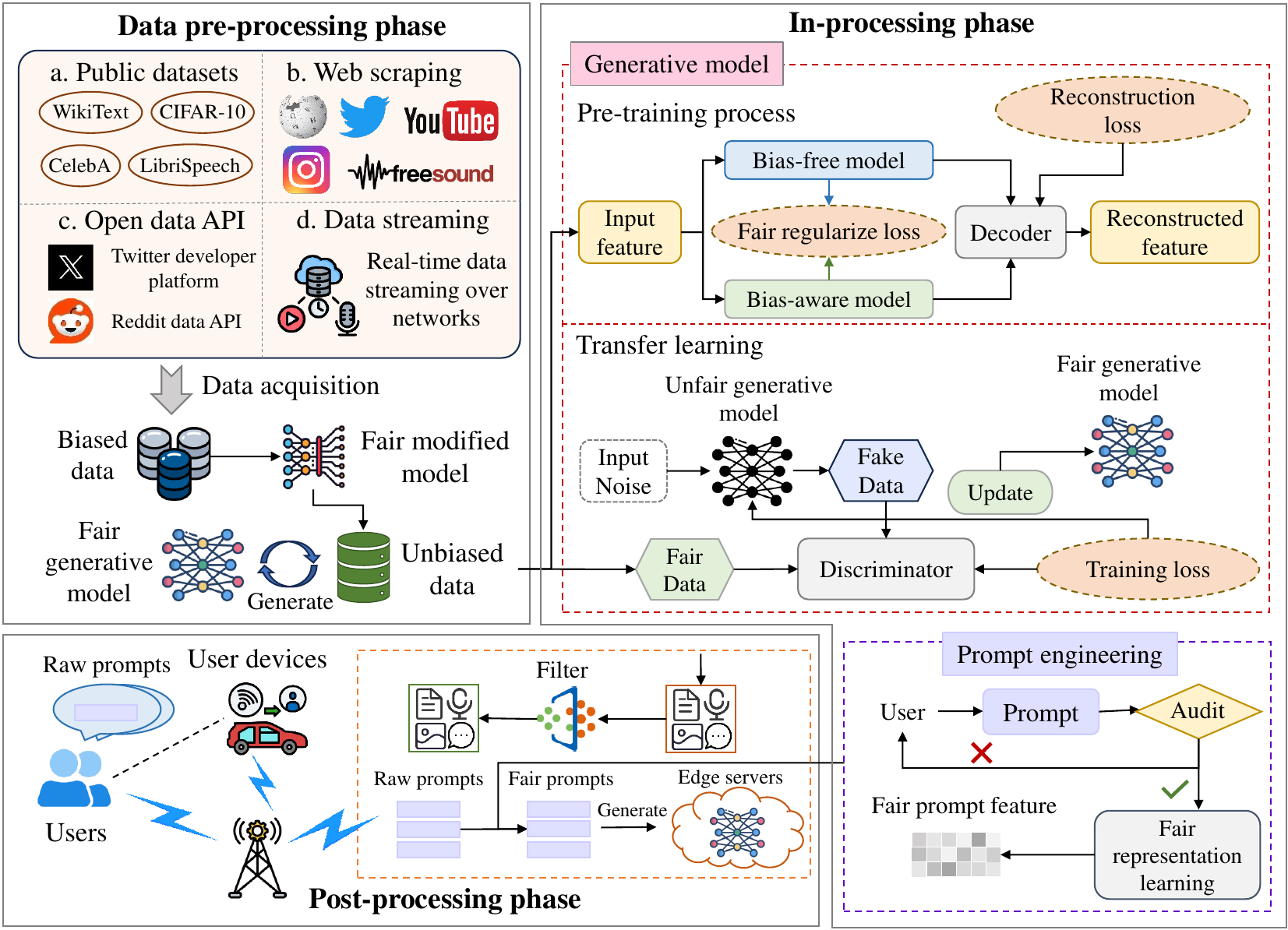}
    \caption{Trustworthy AIGC for achieving more fair network services. 
    The specific methods are divided into three phases: pre-processing, in-processing, and post-processing.
    }
    \label{figure:fairness}
\end{figure*}

\section{Fairness of AIGC for Intelligent Network}
As AIGC models become more prevalent, addressing fairness in generated content is crucial to avoid perpetuating social inequalities, particularly about occupational, gender, and racial stereotypes.
Tackling these biases is key to creating a fair and inclusive network ecosystem that benefits all users.


\subsection{Overview of Unfair AIGC Network Services}
The high data transmission rates and bandwidth of future networks highlight the need to manage large datasets efficiently. 
However, biases in these datasets can lead AIGC models to produce biased outputs. 
For instance, models like ChatGPT and DALL·E exhibit demographic and occupational biases, reflecting the biased representation of groups in the input data. 
This can amplify social biases, especially in sensitive areas like news generation, where underrepresented groups may face discrimination~\cite{fang2024bias}. 
As AIGC technologies grow, biased content could worsen social inequalities.
Since these models are data-driven, they tend to reflect and amplify biases in training data.
Addressing fairness requires a multifaceted approach across pre-processing, in-processing, and post-processing stages, each targeting different biases. 
As shown in \autoref{figure:fairness}, these stages should be applied across fields like content generation, healthcare, and social media to ensure fairness. We examine these stages and discuss measures to enhance fairness in AIGC systems.

\subsection{Methods in Pre-processing Phase}
As shown in the top left of \autoref{figure:fairness}, pre-processing methods address biases in datasets by ensuring fair AIGC outputs.
Datasets used to train generative models often suffer from over-representation or under-representation of certain demographic groups, which can skew results.
To mitigate these biases, independent methods can remove unfair samples, while AIGC itself can generate fair data or augment datasets with more balanced representations. 
For example, image augmentation techniques can adjust sensitive attributes like race or age by manipulating the model’s representation space~\cite{d2024improving}. 
This approach reduces biases without the resource-intensive need for data collection, but metrics for assessing fairness across various applications, such as healthcare or social media, remain under development.

\subsection{Methods in In-processing Phase}
In-processing methods aim to reduce biases during model training or inference. 
These techniques adjust model architecture to promote fairness, either through fairness layers during pre-training or by refining models during transfer learning. 
By incorporating fairness-promoting constraints and using adversarial discriminators, models can be fine-tuned to reduce biases in applications like news generation or e-commerce recommendations~\cite{teo2023fair}. 
Fair representation learning also helps mitigate prompt biases. 
These methods are crucial in ensuring that outputs from generative models are equitable across different sectors, particularly when biases in prompts may otherwise skew the results in sensitive areas like job recommendations or personalized content.
These approaches are illustrated in the in-processing phase of \autoref{figure:fairness}.

\subsection{Methods in Post-processing Phase}
Post-processing techniques focus on filtering biased content in network environments, ensuring fair distribution of generated content across users, as shown in \autoref{figure:fairness}.
In future network architectures, which integrate terrestrial, aerial, and satellite networks, rapid content dissemination can amplify biases. 
To address this, AIGC models are equipped with integrated filters to prevent toxic or biased content. 
However, these filters need further evaluation, as certain methods may inadvertently disadvantage marginalized groups in applications such as social media or customer service. 
Standardizing content filters and establishing fairness metrics is essential to ensure their effectiveness in filtering biases and resisting adversarial attacks, which can bypass moderation systems.

\begin{table}[!t]
    \centering
    \footnotesize
    \caption{Detection scores on fake news, code, and reviews.}
    \begin{tabular}{cccccc} 
    \toprule
    \textbf{Method} & Fake News & Code & Review & \textbf{Avg} \\
    \midrule
    LogRank-based & 53.19 & 62.50 & 52.83 & 56.17 \\
    Entropy-based & 26.48 & 36.09 & 31.52 & 31.36 \\
    RoBERTa detector & 51.46 & 46.26 & 42.47 & 46.73 \\
    GPT Zero-shot & 46.85 & 62.38 & 62.68 & 57.30 \\
    GPT-Zero detector & 59.38 & 69.64 & 70.44 & 66.49 \\
    DetectGPT detector & 45.60 & 69.42 & 71.57 & 62.20\\
    SAD (Ours) & 61.52 & 85.48 & 79.59 & 75.53 \\
    SAD (10\% Paraphrased) & 56.91 & 80.91 & 68.49 & 68.77 \\
    \bottomrule
    \end{tabular}
    \label{table}
\end{table}

\begin{figure}[!t]
    \centering
    \includegraphics[width=\linewidth]{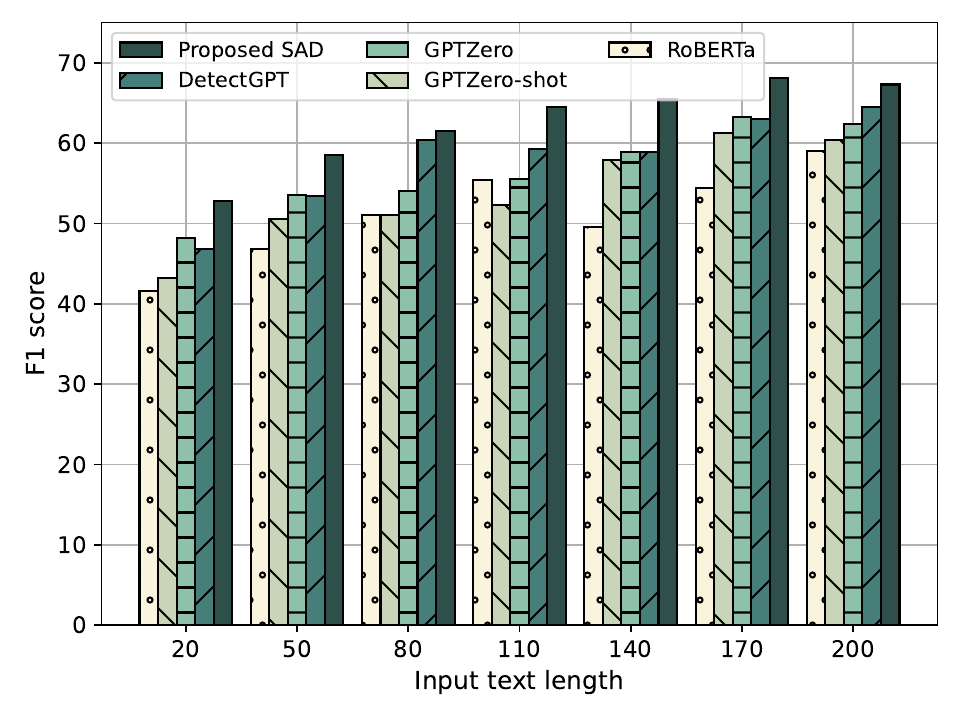}
    \caption{The detection performance of GPT-generated fake news under different input lengths.}
    \label{figure:exp}
\end{figure}

\section{Case Study}
This section presents a case study on adversarial scenarios where LLM-based AIGC models are deployed in dynamic environments that may be vulnerable to malicious attacks. 
In such environments, fluctuating user requests can be mixed with malicious inputs designed to exploit model weaknesses. 
While LLMs like ChatGPT are known for their ability to generate coherent and contextually relevant content, these capabilities can be exploited to carry out cyberattacks or spread false information.
As a case method, we propose SAD, a sentiment analysis-based detector for identifying LLM-generated false information or malicious content in network services.
SAD leverages the emotional intensity difference to detect LLM-generated text, focusing on the unique language feature that occurs when LLMs generate text compared to human-written text.
The approach is based on the premise that LLM-generated content tends to exhibit lower emotional intensity and more consistency, especially when rewritten to neutralize emotional tone.
We designed an analysis pipeline where the input text is rewritten to become more objective, and sentiment analysis is performed on the rewritten text to detect emotional intensity. 
For this task, we use \textit{GPT-3.5-turbo} as the sentiment analysis model, which helps us evaluate and measure the emotional intensity of the text before and after rewriting.
This method allows users to detect subtle emotional discrepancies that are characteristics of LLM-generated texts.

\textbf{Experimental setup.} To benchmark the effectiveness of our approach, we compare TrustGAIN with several state-of-the-art baselines on detecting generated content from \textit{GPT-3.5-turbo} and \textit{GPT-4}. 
GPTZero\footnote{https://gptzero.me/.} is an AI text detection service known as the gold detection standard, capable of detecting \textit{GPT-3.5-turbo} and \textit{GPT-4}.
DetectGPT\cite{mitchell2023detectgpt} is a zero-shot SOTA detector for GPT models, using probability curves as the detection mechanism. 
We use \textit{GPT-3.5-turbo} as representative of current LLMs. 
We employ three datasets to represent the key application scenarios for LLMs. 
The News dataset consists of 500 news articles generated by \textit{text-davinci-003} and 500 real news articles published online. 
The Code dataset includes 500 code blocks from text-davinci-003, such as functions, document strings, and test codes. 
The Review dataset, collected from the Yelp review dataset, consists of 1000 human reviews and 1000 AI-generated reviews. 
All experiments were carried out using PyTorch on two NVIDIA A40 with 48 GB of RAM.

\textbf{Experimental results.}
We conduct detection experiments on three types of unsafe AIGC content: fake news, generated malicious code, and unsafe reviews.
As shown in \autoref{table}, the SAD method outperforms advanced detectors like DetectGPT, GPT Zero-shot, GPT-Zero, and other baselines in identifying unsafe content.
This demonstrates that SAD excels in detecting LLM-generated text, offering a more reliable solution for identifying malicious content in dynamic environments. 
Its consistent performance highlights its effectiveness in addressing the challenges of adversarial LLM-generated content.

Due to the sensitivity of AI-generated text detection to the input length, we evaluate the performance under different input lengths.
The length of the input text to be detected is sampled within the range of 20 to 200, as shown in \autoref{figure:exp}.
The results show that SAD performs better with longer inputs. 
Longer texts provide more context, allowing the model to detect subtle emotional intensity differences that distinguish LLM-generated content from human-written text. 
The model performs better with longer texts, which provide more data for detection.
Overall, SAD demonstrates improved classification accuracy with longer input sizes, proving its robustness in handling varying text lengths while identifying unsafe AIGC content. 
LLM-generated content tends to exhibit a neutral, consistent tone, which becomes easier to identify in longer inputs.
To avoid only discussing a single attack scenario, we illustrate the robustness of SAD to 10\% paraphrasing attacks in \autoref{table}, demonstrating that SAD can still achieve high performance under paraphrasing attacks.
SAD effectively identifies emotional intensity differences in fake news, malicious code, and reviews, with LLM-generated content showing less emotional variation than human-written articles, making it easier to detect. 
This highlights emotional intensity discrepancies as a key feature for detecting LLM-generated content, making SAD valuable for real-world applications with diverse text types and lengths.

\section{Challenges and Future Research Directions}
\textbf{Controversial definition of trustworthy AIGC.} 
The ambiguous definition of trustworthy AIGC affects intelligent network systems, particularly datasets for AI model training. 
Trustworthiness includes robustness, security, fairness, transparency, and privacy, each varying by context.
Fairness encompasses aspects like gender, age, and geography, defined differently across applications.
This lack of a unified definition complicates the standardization of fairness metrics, affecting tasks like dataset labeling in content moderation models, leading to biased decisions.
The diversity of fairness approaches and the absence of consensus remain challenges in designing equitable AI systems. A common framework for trustworthy AIGC is a critical research direction.


\textbf{Challenges of fairness in AIGC network services.}
Advanced AIGC services, reliant on sophisticated algorithms and data infrastructure, are more accessible in economically advanced regions and to individuals with higher technological literacy, exacerbating the digital divide.
This leads to unequal access to AIGC benefits, like productivity enhancement and improved information access.
Biases in training data or models can reinforce stereotypes, disproportionately affecting marginalized communities.
Ensuring fairness in AIGC requires models that consider linguistic, cultural, and socio-economic diversity, along with equitable distribution of resources.
Balancing technical challenges with ethical considerations is crucial to fostering inclusive AI ecosystems, requiring collaboration across research, policy, and engineering to ensure all benefit from AI in future networked environments.


\section{Conclusion}
In this article, we explore the potential of TrustGAIN, a novel AI-native network architecture, in enhancing the credibility of AIGC within intelligent network environments.
TrustGAIN integrates mechanisms for robustness against adversarial threats, security safeguards, and fairness to address the significant ethical challenges associated with AIGC deployment. 
Through preliminary testing for the real datasets in the network environments, we demonstrate that TrustGAIN effectively supports the robust detection of malicious or inaccurately generated fake or malicious content by leveraging advanced LLMs. 
Compared to existing network solutions, the TrustGAIN architecture offers superior capabilities by ensuring more secure, safe, and equitable AIGC services.
Consequently, TrustGAIN represents a crucial advancement toward realizing the full potential of intelligent networks in delivering reliable and fair AIGC services.


\begingroup
\footnotesize
\bibliographystyle{IEEEtran}
\bibliography{main}
\endgroup



\begin{IEEEbiographynophoto}
{Siyuan Li} (siyuanli@sjtu.edu.cn) received his B.Eng. degree from the School of Cyber Science and Engineering, Shanghai Jiao Tong University, Shanghai, China, in 2022, where he is currently a Ph.D. student with the School of Electronic Information and Electrical Engineering.
\\
\\\textbf{Xi Lin} (linxi234@sjtu.edu.cn) is currently an Assistant Professor with the School of Electronic Information and Electrical Engineering, Shanghai Jiao Tong University.
He received his Ph.D. degree from Shanghai Jiao Tong University in 2021 and his B.S. degree from Tianjin University in 2016.
\\
\\\textbf{Yaju Liu} (liuyj7@sjtu.edu.cn) is currently an M.S. student with School of Electronic Information and Electrical Engineering, Shanghai Jiao Tong University, Shanghai, China. \\
\\\textbf{Xiang Chen} (wasdnsxchen@gmail.com) is currently a Ph.D. student with College of Computer Science and Technology, Zhejiang University. \\
\\\textbf{Jianhua Li} (lijh888@sjtu.edu.cn) is a Distinguished Professor and the Dean of Institute of Cyber Science and Technology, Shanghai Jiao Tong University, Shanghai, China. 
He is also the Director of National Engineering Laboratory for Information Content Analysis Technology.
\end{IEEEbiographynophoto}


\end{document}